\documentclass[runningheads,fleqn]{svmult}

\usepackage{epsfig,rotate}
\usepackage{makeidx}   
\usepackage{graphicx}  
\usepackage{subeqnar}  
\usepackage{multicol}  
\usepackage{physmult}  
\makeindex             

\newcommand{\bb}{$\beta\beta~$}          

\begin{document}
\title*{Double Beta Decay: Theory, Experiment, and Implications}
\toctitle{Focusing of a Parallel Beam to Form a Point
\protect\newline in the Particle Deflection Plane}
%
%
\titlerunning{Double beta decay}
%
\author{Petr Vogel}

\authorrunning{Petr Vogel}
%
%

\maketitle              

\section{Introduction}
Double beta decay is a  rare spontaneous nuclear transition 
in which the nuclear charge changes
by two units while the mass number remains the same. 
It has been long recognized as a
powerful tool to study lepton number conservation in general, and
neutrino properties in particular.  
Since the lifetimes of double beta decays are so long, 
the experiments on double beta decay are very challenging and have led
to the development of many generally
valuable techniques to achieve extremely low backgrounds.

For the \bb decay to proceed, the initial nucleus must be less bound 
than the final one, but more bound than the intermediate nucleus.
These conditions are realized in nature for a number of even-even
nuclei (and never for nuclei with an odd number of protons or
neutrons).
Since the lifetime of the \bb decay is always much longer
that the age of the Universe, both the initial and final
nuclei exist in nature 
(some of the actinides being the only exceptions).
In many of the ``candidates'' the transition of two neutrons
into two protons is energetically possible, with the largest $Q$
value  just above 4 MeV. In a few cases the opposite transition,
which decreases the nuclear charge, is also possible,
but the $Q$ values are typically smaller.

The nuclear \bb transition can proceed in several  ways.
One of them, the $2\nu$ decay
\begin{equation}
(Z,A)  \rightarrow (Z+2,A) + e_1^- + e_2^- + \bar{\nu}_{e_1}  + \bar{\nu}_{e_2}
\label{e:2nu}
\end{equation}
conserves the lepton number, while the other one, the
$0\nu$ decay 
\begin{equation}
(Z,A)  \rightarrow (Z+2,A) + e_1^- + e_2^- 
\label{e:0nu}
\end{equation}
violates lepton number conservation and is therefore forbidden
in the standard electroweak theory. 
The prospect of discovering this neutrinoless double beta
decay mode is the driving
force of most of the interest in this field. It is the 
possible window into physics
``beyond the Standard Model''. 

Double beta decay has been  and continues to be a popular topic
since first discussed by Maria Goeppert-Meyer in nineteen thirties.
There have been numerous earlier reviews, beginning with the ``classics'' by
Primakoff and Rosen \cite{Prim}, Haxton and
Stephenson \cite{HS} and Doi, Kotani and Takasugi \cite{DKT},
to the more recent ones, often devoted to particular aspects
of the \bb decay 
\cite{MV94,TZ95,SC98,SPF98,Mo99,Kla99,Ver99}.  Many details 
are also described in the monograph \cite{BV92}.
The Review of Particle Physics \cite{RPP98} regularly summarizes 
the most recent experimental data.

Double beta decay in all its modes is  a  second order weak semileptonic
process, hence its lifetime, proportional to 
($G_F \cos\theta_C$)$^{-4}$, is so very long.  
(Here $G_F = 1.166\times 10^{-5}$GeV$^{-2}$
is the Fermi coupling constant, and $\theta_C$ is the Cabbibo angle.)
The neutrinoless decay can be mediated by a variety of virtual particles,
in particular by the exchange of light or heavy Majorana neutrinos.
The decay amplitude then depends on the masses and coupling
constants of these virtual particles. Independent of the actual mechanism
of the $0\nu$ \bb decay, its observation would imply that neutrinos
necessarily have a nonvanishing Majorana mass \cite{SW82}.
In fact, if the $0\nu$ decay  is actually observed,
and its rate measured, one can obtain, at least in principle,
a lower limit on that mass \cite{Kayser91}.

However, so far no $0\nu$ decay has  been observed. This means
(barring artificial complete cancellation of the amplitudes 
which we dismiss as
unnatural) that the upper limit of the decay rate
can be interpreted as an independent limit for each of the possible 
amplitudes of the decay. In particular, we can obtain the limit on the
properties of light and heavy virtual Majorana neutrinos. Below
we concentrate on the decays mediated by these particles.
(Other possibilities, e.g., the decays mediated
by the new particles predicted by supersymmetry,
are discussed in \cite{Kla99,Ver99}.)

The \bb decay with the Majoron emission, $0\nu\chi$ mode,
\begin{equation}
(Z,A)  \rightarrow (Z+2,A) + e_1^- + e_2^-  + \chi ~.
\label{e:mnu}
\end{equation}
belongs to the category of the lepton number violating decays,
even though the lepton number is formally conserved when
$\chi$ is assigned the lepton number $-2$. The hypothetical scalar particle
$\chi$, which must be in this case light enough to be emitted in the \bb
decay,  is usually associated with spontaneous 
breaking of the $B-L$ symmetry \cite{Chi80,GR81}.

\begin{figure}[h]
\includegraphics[width=.8\textwidth]{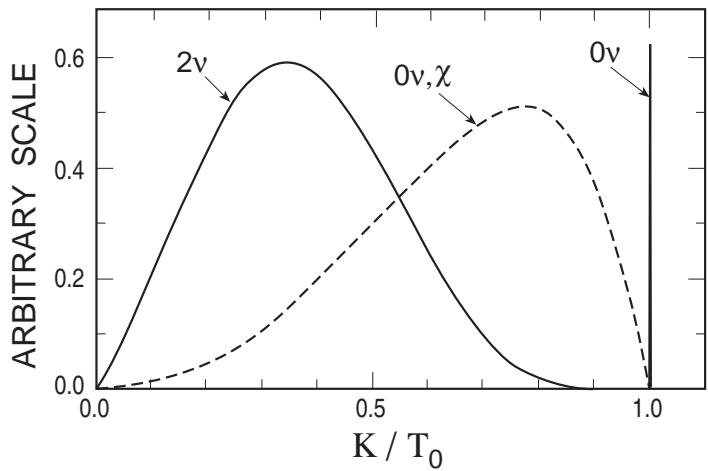}
\caption[]{Schematic sum electron spectra of the three \bb
decay modes. Each is normalized arbitrarily and independently of
the others. The abscissa is the ratio $K/T_0$ of the sum-electron kinetic energy
divided by its maximum value.}
\label{fig:spect}
\end{figure}

Empirically, it is easy to distinguish between the three decay modes 
listed above,
provided the electron energies are measured. The electron sum energy spectra
are determined by the phase space of the outgoing leptons
and clearly characterize the decay mode, as schematically
illustrated in Figure \ref{fig:spect}. 
(Geochemical or milking experiments, however, cannot distinguish between
the different \bb modes as they determine only the total decay rate.)

There are two distinct groups of theoretical issues associated with
the interpretation of the \bb decay experiments. The particle physics issues
deal with the expression of the decay rate in terms of the fundamental
parameters, such as the neutrino masses and mixing angles, 
coupling constants in the weak interaction Hamiltonian, etc. 
This group of problems involves also
the relation of the \bb decay to other processes, such as neutrino oscillations,
direct mass measurements, and searches for other lepton number violating 
processes. 

The other, essentially decoupled, set of problems involves the nuclear structure
issues associated with the \bb decay. The decay rate is expressed in terms
of nuclear matrix elements (NME) which have to be evaluated. One would like
to know, first of all, their value and its uncertainty. This area
of research has attracted lots of attention, and there are many, often
conflicting, evaluations available in the literature. Unfortunately,
there is no simple way of judging the correctness and accuracy of the 
evaluations of the nuclear matrix elements for the neutrinoless decay.
Comparison to the experimentally known rate of
the $2\nu$ \bb decay rate is often invoked in that context
as a test of the ability of the nuclear model to describe the
related phenomena. It is not clear, however, if this is indeed
a valid test. For example,
if one assumes that the $0\nu$ decay is mediated by the
exchange of a heavy particle (whether this exchanged particle
is a heavy neutrino or not), the corresponding internucleon
potential is of short range, and additional issues involving nucleon
structure, irrelevant for the $2\nu$ decay, play an important role.
Another, less fundamental but in praxis perhaps more important
example deals with the dependence of the NME on the number of
single nucleon subshells included in the calculation. For
the $2\nu$ decay, where only the Gamow-Teller operator $\sigma\tau$
plays a role, it is clearly sufficient to include just the states
within the valence oscillator shell. It is less clear  that the
same truncation is sufficient for the  correct description
of  the $0\nu$ decay.

The experimental study of the \bb decay presents a formidable challenge
since the goal is to detect a process with a half-life in excess
of $10^{25}$ years (the present best limit for the $0\nu$ decay). 
The \bb decay must be detected in presence of an inevitable
background of a similar energy caused by 
trace radioisotopes with half-lives 15 or more orders of magnitude
shorter. Thus, the optimum separation of the signal from background, combined
with the requirement of having kilogram quantities of the source isotopes,
characterizes the present day experiments. 

The past and current experiments are still relatively modest in size,
and therefore also in complexity and cost.
Given the importance of the search for the neutrinoless decay, ambitious plans,
involving much larger amounts of the source nuclei, are considered.
Naturally, the larger source mass will be beneficial 
only if it is accompanied by the corresponding
reduction of the background. The future projects will be therefore inevitably much more
complex,  and will involve larger groups of researchers. With them,
the field of \bb decay, which competes already with the other experiments 
described in this book in importance, 
will also compete  in size and cost.

\section{Lepton number violation}
With the usual assignment of the lepton number,
$L(l^-) = L(\nu) = -L(l^+) = -L(\bar{\nu}) = +1$, 
$0\nu$ \bb decay  represents a change
in  the global lepton number by two units, $\Delta L = 2$. 
In that respect its observation would be related to the attempts 
to detect $\bar{\nu}_e$ from the sun, or of $\nu_e$ from nuclear 
reactors. Both of these latter processes represent a kind of
``$\nu \leftrightarrow \bar{\nu}$ oscillations'',
and also are possible only for massive Majorana neutrinos.

For light Majorana neutrinos the lepton number conservation is irrelevant
and the $0\nu$ decay is hindered only by the helicity mismatch. However,
the ``antineutrino'' born in association with one of the $e^-$ in the $0\nu$
decay is not fully
righthanded, but has a lefthanded component of amplitude $\sim m_{\nu}/E_{\nu}$.
This lefthanded piece can be absorbed by another neutron which is
converted into a proton and the second $e^-$ is emitted.  Similar consideration
would govern the above mentioned $\nu \leftrightarrow \bar{\nu}$ oscillations.
The word ``oscillations'' in this context is a misnomer,
however, since the process (if it exists)
would proceed without an oscillatory behavior \cite{KFM99}.

The expected branching ratio
for the ``wrong'' neutrinos at low energies, relevant for the sun or
nuclear reactors is \cite{LW98}
\begin{equation}
R \sim \frac{m_{\nu}^2}{2 E_{\nu}^2} 
\frac{ \sigma^{\bar{\nu}N}}{ \sigma^{\nu N}} ~\sim~ 10^{-14} ~,
\end{equation}
where the numerical factor was derived for $m_{\nu} \sim 1$ eV,
$E_{\nu} \sim 5$ MeV, and the ratio of  cross sections put to unity.
Since the $0\nu$ \bb decay is presently sensitive to such neutrino
masses, one cannot expect a signal for this kind of 
$\nu \leftrightarrow \bar{\nu}$ oscillations until similar sensitivity
is achieved, i.e., not anytime soon, if ever.

However, it is also possible that $\bar{\nu}_e$ from the sun
are produced in a more complicated, but possibly more 
efficient way. Let us assume that a transition magnetic moment $\mu_{e,l}$
connects the lefthanded $\nu_{e,L}$ with a righthanded $\bar{\nu}_{l,R}$
of a different flavor, which can subsequently oscillate (by the vacuum
or matter enhanced oscillations) into the righthanded and thus
observable $\bar{\nu}_{e,R}$, i.e.,
when neutrinos propagate in a transverse 
solar magnetic field $B_{\perp}$
one or both of the sequences 
$\nu_{e,L} \rightarrow \bar{\nu}_{l,R} \rightarrow \bar{\nu}_{e,R}$ or
$\nu_{eL} \rightarrow \nu_{l,L} \rightarrow \bar{\nu}_{eR}$
occurs. Such process requires that the magnetic conversion,
which is possible only for the massive Majorana neutrinos and
which depends on the product $\mu_{e,l}B_{\perp}$, and the flavor oscillation,
which depends on $\Delta m^2$ and sin$^2 2\theta$, are both 
present. There is no obvious relation between this process and the 
neutrinoless \bb decay, except that both require the existence of the
neutrino Majorana mass term.
(This brief discussion of the magnetic conversion is highly simplified.
In reality, the transition magnetic moments ought to be written
in terms of mass eigenstates \cite{BV99}.)

Finally, tight experimental limits exist on the total lepton number violating 
processes which involve both electrons and muons (see \cite{RPP98}),
such as the muon conversion
\begin{equation}
\mu^- + (Z,A) \rightarrow (Z-2,A) + e^+ ~,
\end{equation}
and the muonium-antimuonium conversion
\begin{equation}
\mu^+ ~  e^- \rightarrow \mu^- ~ e^+ ~.
\end{equation}
The relation of these processes to the
\bb decay is, however, not well established.

\section{Particle physics aspects}

In this section we shall consider how the rate 
of the neutrinoless \bb decay is related to the
unknown parameters of the  neutrino mass matrix
and to the phenomenological parameters describing
a generalized semileptonic charged 
current weak interactions $H_W$:

\begin{equation}
H_W ~=~ \frac{G_F}{\sqrt{2}} \left[ 
J_L^{\alpha} ( M_{L\alpha}^+ + \kappa M_{R\alpha}^+ ) ~+~ 
J_R^{ \alpha} (\eta M_{L\alpha}^+ + \lambda M_{R\alpha}^+ )
\right] ~+~  \rm H.c. ~,
\label{eq:ham}
\end{equation}
where $J_{L(R)}$ and $M_{L(R)}$ are the lepton and quark 
left(right)-handed current four-vectors, respectively. The 
dimensionless parameters
$\eta, \lambda,$ and $\kappa$ characterize deviations from the
standard model. (Since $\kappa$ gives a negligible contribution
to double beta decay, we will not consider it from now on.) 
The coupling parameters $\eta$ and $\lambda$, modified by the
neutrino mixing, and denoted then usually as
$\langle \eta \rangle$ and $\langle \lambda \rangle$ are unknown
(and presumably small).

The lepton sector of the theory 
contains in general $n$ generations
of charged leptons as well as $n$ left- and $n$ right-handed neutrinos.
The neutrino mass matrix is the $2n \times 2n$ matrix $M$
\begin{equation}
 M = \left( \begin{array}{cc}
               M_L    &  M_{D} ^T \\
	       M_D    &  M_R
	       \end{array}    \right) ~,
\label{eq:mass}
\end{equation}	       
where $M_D$ is the $n \times n$ lepton number conserving Dirac mass term,
and the symmetric $n \times n$ matrices $M_L$ and $M_R$ are the lepton
number violating Majorana mass terms. The matrix $M$ has $2n$ real, but
not necessarily positive, eigenvalues.
Writing the eigenvalues as $m_j \epsilon_j$, 
we can impose the physically reasonable
condition that $m_j \geq 0$. The sign of the eigenvalues of the mass
matrix is contained in the phases $\epsilon_j = \pm 1$ 
which are the
intrinsic $CP$ parities of the neutrinos $j$.

Neutrino oscillation phenomena arise
because the ``mass eigenstates" of $M$ or, 
more precisely their chiral projections
$N_j^L$ and $N_j^R$, are not necessarily the 
familiar weak interaction neutrinos that couple
to the known intermediate vector boson $W_L$ and to the hypothetical
right-handed boson $W_R$.   The physical ``weak eigenstate" 
or current
neutrinos,  the $n$  left-handed neutrinos $\nu_L$  and the $n$
right-handed ones $\nu_R^{'}$ (the prime has been added
in order to stress that they are {\it different} particles),
are related to the neutrinos of definite mass by the $n \times 2n$ mixing
matrices $U$ and $V$
\begin{equation}
\nu_L ~=~ U N^L ~~, ~~ \nu_R^{'}  ~=~ V N^R ~.
\label{eq:mix}
\end{equation}
The mixing matrices $U$ and $V$ obey the  normalization and
orthogonality conditions
\begin{equation}
\sum_{j=1}^{2n} U_{l j}^{*} U_{l' j} ~=~ \delta_{l l'} ~,~
\sum_{j=1}^{2n} V_{l j}^{*} V_{l' j} ~=~ \delta_{l l'} ~,~
\sum_{j=1}^{2n} U_{l j}^{*} V_{l' j} ~=~ 0 ~.
\label{eq:ortho}
\end{equation}

In neutrinoless \bb decay the rate depends
on the effective parameters which are expressed
in terms of the mixing matrices $U$ and $V$:
\begin{eqnarray}
\label{eq:eff}
\langle m_{\nu} \rangle ~ & = & ~ \sum_j  {^{'}}
\epsilon_j m_j U_{e,j}^2 ~,  \nonumber \\
\langle \lambda \rangle ~ & = & ~ 
\lambda \sum_j {^{'}} \epsilon_j  U_{e,j} V_{e,j} ~,   \\
\langle \eta \rangle ~ & = & ~ 
\eta \sum_j {^{'}} \epsilon_j  U_{e,j} V_{e,j} ~,  \nonumber \\
\langle g_{\nu,\chi} \rangle ~ & = & ~  
\frac{1}{2} \sum_{i,j} {^{'}}  (g_{i,j}\epsilon_i + g_{j,i}\epsilon_j)~ 
U_{e,i} U_{e,j} ~.  \nonumber 
\end{eqnarray}
Here the prime indicates that the summation is over only 
relatively light neutrinos. Also, 
$\lambda$ and $\eta$ are the dimensionless coupling constants for the
right-handed current weak interaction, 
Eq.(\ref{eq:ham}), and $g_{i,j}$ are the
coupling constants of interaction between the Majoron $\chi$ and the
Majorana neutrinos $N_i$ and $N_j$.
For the heavy neutrino one obtains
\begin{equation}
\langle m_{\nu}^{-1} \rangle_H ~ = ~ \sum_j  {^{''}}
\epsilon_j m_j^{-1} U_{e,j}^2 ~,  
\label{eq:heavy}
\end{equation}
where the double prime indicates that the summation, involving
the inverse neutrino masses $ m_j^{-1}$, is over only 
the heavy neutrino mass eigenstates ($m_j \ge $1 GeV).

It is now clear that, within the mechanism considered
so far, there is no neutrinoless double beta 
decay if all neutrinos
are massless. Not only $\langle m_{\nu} \rangle$ 
vanishes in such a case but also
$\langle \lambda \rangle$ and $\langle \eta \rangle$  vanish
due to the orthogonality condition Eq. (\ref{eq:ortho}). Moreover, 
$\langle \lambda \rangle$ and $\langle \eta \rangle$ vanish for the same
reason even if some or all neutrinos are massive but light 
and therefore the summation in Eq.(\ref{eq:eff}) 
contains all neutrino mass eigenstates.
In that case, however, there is a smaller next order contribution from
the mass dependence of the neutrino propagator,
which for this purpose can be written as
\begin{equation}
\frac{\gamma_{\mu} q^{\mu}}{q^2 + m_j^2}  ~\approx~
\frac{\gamma_{\mu} q^{\mu}}{q^2} 
\left( 1 - \frac{m_j^2}{q^2} \right) ~.
\end{equation}
The expression for e.g., $\langle \lambda \rangle $ now
contains $\sum_j {^{'}} \epsilon_j  U_{e,j} V_{e,j} m_j^2$
which clearly shows that a nonvanishing neutrino mass
is required. 

The presence of the phases $\epsilon_j$ in the expression 
for $\langle m_{\nu} \rangle$
means that cancellations are possible. In particular, 
for every Dirac neutrino there is
an exact cancellation, since the Dirac neutrino is equivalent to a pair of
Majorana neutrinos with the opposite sign of 
the phases $\epsilon_j$ and degenerate masses.

In the general case the neutrinoless double beta decay rate is a
quadratic polynomial in the unknown parameters
\begin{eqnarray}
[ T_{1/2}^{0\nu} (0^+ \rightarrow 0^+)]^{-1} & = &
C_1\frac{\langle m_{\nu} \rangle^2}{m_e^2} + C_2\langle \lambda \rangle
\frac{\langle m_{\nu} \rangle}{m_e}\cos\psi_1 
+  C_3 \langle \eta \rangle 
\frac{\langle m_{\nu} \rangle}{m_e}\cos\psi_2 \nonumber \\
& + & C_4 \langle \lambda \rangle^2 
+ C_5 \langle \eta \rangle^2 
+ C_6 \langle \lambda \rangle \langle \eta \rangle \cos(\psi_1 - \psi_2) ~.
\label{eq:tott}
\end{eqnarray}
Here $\psi_1$ and $\psi_2$ are the phase angles between the generally
complex numbers $m_{\nu}$, $\lambda$ and $\eta$. (However, when 
$CP$ invariance is assumed $\psi_{1,2}$ are either 0 or $\pi$.) 
The phase space integrals
{\em and} the nuclear matrix elements are combined in the
factors $C_i$. 
Assuming that we can calculate them, Eq. 
(\ref{eq:tott}) represents an ellipsoid which restricts the 
allowed range of
the unknown parameters 
$\langle m_{\nu} \rangle$, $\langle \lambda \rangle$ and
$\langle \eta \rangle$ for a given value (or limit) of the
$0\nu$ double beta decay lifetime.

In order to evaluate the nuclear matrix elements,
we must consider the neutrino propagator. 
Assuming that $\langle m_{\nu} \rangle^2$
is the only relevant quantity, one can perform
the integration over the four-momentum of the
exchanged particle and obtain the
``neutrino potential'', which for $m_{\nu} <$ 10 MeV has the form
\begin{equation}
H(r,\Delta E) ~=~ \frac{2R}{\pi r} \int_{0}^{\infty} dq \frac{\sin(qr)}
{q+\Delta E} ~,
\label{eq:nupot}
\end{equation}
where $\Delta E = \langle E_N\rangle - 1/2(M_i+M_f)$ is 
the average excitation energy of the intermediate
odd-odd nucleus and the factor $R$ (the nuclear radius) has been
added to make the neutrino potential dimensionless.

When the $0\nu$ decay is mediated by the right-handed weak current interaction
the evaluation of the decay rate becomes more complicated, 
since many more terms
must be included (see \cite{HS,DKT,Tomoda}). If the four-momentum of the virtual
neutrino is  $q_{\mu} \equiv \omega, \vec{q}$, the neutrino
propagator contains 
\begin{eqnarray}
\omega \gamma_0 - \vec{q} \cdot \vec{\gamma} + m_j  ~.     \nonumber
\end{eqnarray}
The part of the propagator proportional to $m_j$ is responsible for the
neutrino potential Eq.(\ref{eq:nupot}). The part containing $\vec{q}$ leads to
a new potential related to the derivative of $H(r,\Delta E)$, and the 
part with $\omega$ leads to yet another potential, which is a combination
of $H(r,\Delta E)$ and its derivative.

Similarly, there are now also more nuclear matrix elements, which contain
in addition the nucleon momenta (i.e., the gradient operators), and depend on
the nucleon spins and radii in a more complicated way (e.g.,  they contain 
tensor operators). The outgoing electrons are no longer just in the
$s_{1/2}$ states, because for some of the operators one of the electrons will
be in the $p_{1/2}$ state. 
The recoil matrix element, which originates from the recoil term in the
nuclear vector current is
numerically relatively large \cite{Tomoda}, resulting in
more sensitivity to the parameter  $\langle \eta \rangle$.
The current best limits on $\langle \eta \rangle$ and
$\langle \lambda \rangle$ are listed in \cite{RPP98}.

\section{Experimental techniques and results}

It is beyond the scope of this review to describe in detail the experimental techniques
developed to meet the challenge of background suppression
and signal recognition needed to determine the rate (or an interesting
limit) of the \bb decay. Thus, only the briefest outline is given,
and the most important experimental results are summarized in
tables.

Historically, the existence of \bb decay was first established using
the {\it geochemical} method. Here one takes advantage 
of geologic integration times by 
searching for daughter products accumulated in ancient minerals that are rich 
in the parent isotope. (The related {\it radiochemical} method is applicable
if the daughter isotope is radioactive.) Since the energy information 
is long lost, the mode of  \bb
decay responsible is not directly determined. Instead, the total decay rate
is determined, and thus an upper limit of each mode as well.
 
Only by measuring the energies of electrons released in the decay 
in the direct counting experiments can 
one distinguish directly the mode of decay. The $2\nu$ and $0\nu \chi$ decay
modes each result in a rather generic looking 
electron spectrum (see Fig.\ref{fig:spect}),
and the observation of these decays requires either an extremely efficient
background suppression or additional information, such as a tracking capability.

The measured half-lives of the $2\nu$ mode are collected in Table \ref{tab:2nu}. 
Many of them have been measured by several groups; only the results with
the smallest claimed errors are shown. (The case of $^{130}$Te where
the two competing results have the same error but exclude each other is 
the only exception.) Also, the numerous half-life limits have been
omitted. The $2\nu$ mode is now well established; no doubt many more
and more accurate results will become available soon.

In fact, \bb decay is becoming a valuable tool of nuclear spectroscopy.
The decay of $^{100}$Mo into the excited $0^+$ state at 1130 keV
in $^{100}$Ru has been observed \cite{Barab,Ludwig}. The technique used,
the observation of the subsequent $\gamma$ decay cascade, can be 
readily applied to other nuclei as well. This development not only
expands the scope of the experimental study of the \bb decay, but allows more
detailed comparison between theory and experiment
(for an early attempt, see \cite{Austyn}).
 
\begin{table}[htb]
\begin{center} 
\caption{\protect Recent $\beta\beta_{2\nu}$ results.
\newline
(Only positive results are listed. The most accurate
published values are given, except for
$^{130}$Te where two conflicting results with the
same claimed errors are quoted.)}
\label{tab:2nu}
\renewcommand{\arraystretch}{1.2}
\begin{tabular}{llc} \\ \hline\hline
Isotope &T$_{1/2}^{2\nu}$ (y) & Reference \\ \hline
$^{48}$Ca   &$(4.3_{-1.1}^{+2.4} \pm 1.4) \times 10^{19}$  &\cite{Bal96} \\
$^{76}$Ge   &$(1.77 \pm 0.01  ~ _{-0.11}^{+0.13} )  \times 10^{21}$  
 & \cite{Gun97}  \\
$^{82}$Se   &$(8.3 \pm 1.0 \pm 0.7 ) \times 10^{19}$   & \cite{Arn98} \\
$^{96}$Zr   &$(3.9\pm 0.9)\times 10^{19}$ $^{geoch}$ & \cite{Kaw93} \\
$^{100}$Mo  &$(6.82^{+0.38}_{-0.53} \pm 0.68) \times 10^{18}$ 
 & \cite{Desilva97} \\
 $^{116}$Cd  &$(3.75 \pm 0.35 \pm 0.21)\times 10^{19}$    & \cite{Arn96} \\
$^{128}$Te  &$(7.2\pm 0.4)\times 10^{24}$ $^{geoch}$  & \cite{Ber92} \\
$^{130}$Te  &$(2.7\pm 0.1)\times 10^{21}$  $^{geoch}$  & \cite{Ber92} \\
  & $(7.9 \pm 1.0) \times 10^{20}$ $^{geoch}$ & \cite{Tak96} \\
$^{150}$Nd  &$(6.75_{-0.42}^{+0.37} \pm 0.68)\times 10^{18}$  
 & \cite{Desilva97} \\
$^{238}$U   &$(2.0\pm 0.6)\times 10^{21}$ $^{radioch}$  & \cite{Tur91} \\ \hline
\end{tabular}
\end{center}
$^{geoch}$ geochemical determination; total decay rate.
\newline
$^{radioch}$ radiochemical determination; total decay rate
\end{table}

The 0$\nu$ mode can be approached quite differently from 2$\nu$ 
and 0$\nu,\chi$ modes because of the 
distinctive character of the 0$\nu$ electron sum spectrum --- 
a monoenergetic line at the full 
{\it Q}-value (see Fig. \ref{fig:spect}).  Obviously,
sharp energy resolution of 0$\nu$ detectors is a big advantage
which helps to isolate the line from background.  As in the case
of the 2$\nu$ decay other features, such as tracking, naturally
help as well. 

The best reported limits for the neutrinoless \bb decay modes are
collected in Tables \ref{tab:0nu} and \ref{tab:maj}. Again, only the most
restrictive limits for the given transition are shown. The longest half-life
limit, reported for $^{76}$Ge by the Heidelberg-Moscow collaboration
\cite{Baud99}, is based on 24.16 kg yr of exposure and uses pulse
shape discrimination to suppress the background (in the relevant 
energy region the background is a mere ($0.06 \pm 0.02$) 
events/(kg$\cdot$yr$\cdot$keV)).
In that experiment 7 events were observed in the $3\sigma$ region around
the $0\nu$ decay $Q$ value, while from the background extrapolation one
expects 13 events. Using this lack of background events an even more stringent
limit (the entry in parenthesis in Table \ref{tab:0nu}) is obtained. 

The limit based on the Te lifetime ratio in Table \ref{tab:0nu} is based
on the different $Q$ value dependence of the $0\nu$ and $2\nu$ modes.
That this offers a valuable tool has been recognized already in the
prophetic early paper by Pontecorvo \cite{Pont}. Even though
the corresponding NME are not exactly equal, they are close enough
to allow one to use the geochemical lifetime determination here
and in Table \ref{tab:maj}. 

\begin{table}[ht]
\caption{\protect Best reported limits on T$_{1/2}^{0\nu}$ 
and $\langle m_{\nu}\rangle$. 
\newline
The experimental result is listed first with its reference. 
This is followed by the
limit on $\langle m_{\nu}\rangle$ followed by the reference to the employed
nuclear matrix element (NME). Whenever possible 
the choice of the authors of the
experimental paper regarding the NME is respected. See the text
for the discussion of uncertainties associated with the evaluation
of NME's.}
\label{tab:0nu}
\begin{center} 
\renewcommand{\arraystretch}{1.2}
\begin{tabular}{llccc} \\ \hline\hline
Isotope &T$_{1/2}^{0\nu}$ ($10^{22}$ y) (CL\%) & exp. ref.
 &$\langle m_{\nu}\rangle$ (eV)
& NME ref. \\ \hline
$^{48}$Ca          &$>0.95$ (76) &  \cite{You91}         &$<18.3$   & \cite{HS} \\
$^{76}$Ge     &$>1600(5700)^{~(see ~1)}$ (90)    & \cite{Baud99} &$<0.4 (0.2)^{~(see ~1)}$ 
&  \cite{Muto} \\
$^{82}$Se    &$>2.7$ (68) &   \cite{Ell92}           &$<5$ 
& \cite{HS}  \\
$^{100}$Mo         &$>5.2 $ (68)  &   \cite{Eji96}         & $<6.6$
&  \cite{Tomoda} \\
$^{116}$Cd         &$>2.9 $ (90)   &\cite{Geor95}    &$<4.6$ 
& \cite{Muto}  \\
$\frac{T_{1/2}(130)}{T_{1/2}(128)}$ $^{(see ~ 2)}$ &
$(3.52\pm 0.11)\times 10^{-4}$    &  \cite{Ber92}     
&$<1.1-1.5$ &  \cite{Muto,Engel} \\
$^{136}$Xe     &$>44$ (90) & \cite{Leusch98}
&$<2.3-2.8$  &\cite{Engel} \\ 
$^{150}$Nd        &$> 0.12 $ (90) &  \cite{Desilva97}
& $<4.0$  &  \cite{Muto} \\ \hline
\end{tabular}

$^{1}$ the first entry is based on the average background 
and the entry in parenthesis is based on the apparent lack of background counts
in the corresponding energy interval.

$^{2}$ geochemical determination of the lifetime ratio
\end{center}
\end{table}

\begin{table}[ht]
\caption{\protect The most restrictive Majoron limits.}
\label{tab:maj}
\begin{center} 
\renewcommand{\arraystretch}{1.2}
\begin{tabular}{lllc} \\ \hline\hline
Isotope &T$_{1/2}^{0\nu,\chi}$ (y) and (CL \%) &$\langle g_{\nu,\chi}\rangle$ 
& Reference \\ \hline
$^{48}$Ca   &$>7.2 \times 10^{20}$  (90) 
&$<5.3\times 10^{-4}$ & \cite{Bar89b} \\
$^{76}$Ge   &$>1.66\times 10^{22}$ (90) 
&$<1.8\times 10^{-4}$,  & \cite{Beck93} \\
$^{82}$Se   &$>2.4\times 10^{21}$ (68)  
&$<2.3 \times 10^{-4}$  & \cite{Arn98}\\
$^{100}$Mo   &$>5.4\times 10^{21}$ (68) 
& $<7.3 \times 10^{-5}$ & \cite{Eji96} \\
$^{116}$Cd    & $>1.2 \times 10^{21}$ (90) & $<2.1 \times 10^{-4}$
& \cite{Dan98} \\
$^{128}$Te  &$>7.7\times 10^{24 ~geoch}$ (90) & $<3\times 10^{-5}$ 
& \cite{Ber92} \\
$^{136}$Xe  &$>7.2\times 10^{21}$ (90)  
&$<1.6\times 10^{-4}$ &\cite{Leusch98} \\
$^{150}$Nd  &$>2.8\times 10^{20}$ (90) 
& $<1\times 10^{-4}$ &  \cite{Desilva97} \\ \hline
\end{tabular}

$^{geoch}$ geochemical determination; from total decay rate

\end{center}
\end{table}

\section{Nuclear  structure aspects}

The rate of the $2\nu$ \bb decay is simply
\begin{equation}
1/T_{1/2}^{2\nu} = G_{2\nu} (E_0, Z) | M_{2\nu} |^2 ~,
\end{equation}
while for the neutrinoless decay (assuming that it is mediated
by a light Majorana neutrino and that there are no right-handed
weak interactions), and for the decay 
with Majoron emission, it is given by
\begin{eqnarray}
1/T_{1/2}^{0\nu} & = & G_{0\nu} (E_0, Z) | M_{0\nu} |^2 
\langle m_{\nu} \rangle^2 ~, \\ \nonumber
1/T_{1/2}^{0\nu,\chi} & = & G_{0\nu,\chi} (E_0, Z) | M_{0\nu,\chi} |^2
\langle g_{\nu,\chi} \rangle^2 ~.
\end{eqnarray}
Here the phase space functions $G(E_0,Z)$ are accurately calculable,
and the nuclear matrix elements $M$ are the topic of this section.
Obviously, the accuracy with which
the fundamental particle physics
parameters $\langle m_{\nu} \rangle$
and $ \langle g_{\nu,\chi} \rangle$ can be determined is
limited by our ability to evaluate these nuclear matrix elements.

In that context there are three distinct set of problems:

\begin{itemize}
\item $2\nu$ decay: the physics of the Gamow-Teller amplitudes
\item $0\nu$ decay  with the exchange of light massive Majorana neutrinos:
no selection rules on multipoles, role of nucleon correlations, sensitivity
to nuclear models.
\item $0\nu$ decay with the exchange of heavy neutrinos: physics 
of the nucleon-nucleon states at short distances.
\end{itemize}

\subsection{Two neutrino decay} 

Since the energies involved are
modest, the allowed approximation should be applicable, and the rate
is governed by the double Gamow-Teller matrix element 
\begin{equation}
M_{GT}^{2\nu} = \sum_m \frac{\langle f || \sigma \tau_+ || m \rangle
\times  \langle m || \sigma \tau_+ || i \rangle } { E_m - (M_i + M_f)/2 } ~'
\label{eq:2nu}
\end{equation}
where $i, f$ are the ground states in the initial and final nuclei, 
and $m$ are the intermediate $1^+$ (virtual) states in the odd-odd nucleus.
The first factor in the numerator above represents the $\beta^+$
(or ($n,p$)) amplitude for the final nucleus, while the second one
represents the $\beta^-$ (or ($p,n$)) amplitude for the initial nucleus.
Thus, in order to correctly evaluate the $2\nu$ decay rate, we have to
know, at least in principle, {\it all} GT amplitudes for 
both $\beta^-$ and $\beta^+$ processes,
including their signs. The difficulty is that the $2\nu$ matrix element
exhausts a very small fraction ($10^{-5} - 10^{-7}$) of the double GT
sum rule \cite{double}, and hence it is sensitive to details of 
nuclear structure.

Various approaches used in the evaluation of the $2\nu$ decay rate
have been reviewed recently in Ref. \cite{SC98}.
The Quasiparticle Random Phase Approximation (QRPA) has been the most
popular theoretical tool in the recent past. Its main ingredients, the
repulsive particle-hole spin-isospin interaction, and the attractive
particle-particle interaction, clearly play a decisive role in the
concentration of the $\beta^-$ strength in the giant GT resonance,
and the relative suppression of the $\beta^+$ strength and its 
concentration at low excitation energies. Together, these two 
ingredients are able to explain the suppression of the $2\nu$
matrix element when expressed in terms of the corresponding sum rule.

Yet, the QRPA is often criticized. Two ``undesirable'', and to some 
extent unrelated, features are usually quoted.
One is the extreme sensitivity of the decay rate to the
strength of the particle-particle force (often denoted as $g_{pp}$).
This decreases the predictive power of the method. The other one is the
fact that for a realistic value of  $g_{pp}$ the QRPA solutions are close
to their critical value (so called collapse). This indicates a phase
transition, i.e., a rearrangement of the nuclear ground state. QRPA
is meant to describe small deviations from the unperturbed ground state,
and thus is not fully applicable near the point of collapse. 
Numerous approaches have been made to extend the range of validity of QRPA, 
see e.g. \cite{SC98}. Altogether, QRPA and its various 
extensions, with their ability to adjust at least one free parameter,
are typically able to explain the observed $2\nu$ decay rates.

At the same time, detailed calculations show that 
the sum over the excited states in Eq.(\ref{eq:2nu}) converges 
quite rapidly \cite{EEV94}. In fact, a few low lying states usually
exhaust the whole matrix element. Thus, it is not really necessary to describe
all GT amplitudes; it is enough to describe correctly the $\beta^+$ and
$\beta^-$ amplitudes of the low-lying states, and include everything else in
the overall renormalization (quenching) of the GT strength. 

Nuclear shell model methods are presently capable of handling much larger
configuration spaces than even a few years ago. Thus, for many nuclei the
evaluation of the $2\nu$ rates within the 0$\hbar \omega$ shell model
space is feasible. (Heavy nuclei with permanent deformation, like
$^{150}$Nd and $^{238}$U remain, however, beyond reach of the
shell model techniques.)
Using the shell model avoids, naturally, the above difficulties of QRPA.
At the same time, the shell model can describe, using the same method
and the same residual interaction, a wealth of spectroscopic data,
allowing much better tests of its predictive power.

\subsection{Neutrinoless decay: light Majorana neutrino}

If one assumes that the $0\nu$ decay is caused by the exchange
of a virtual light Majorana neutrino between the two nucleons,
then several new features arise: a) the exchanged neutrino
has a momentum $q \sim 1/r_{nn} \simeq 50 - 100$ MeV
($r_{nn}$ is the distance between the decaying nucleons).
Hence, the dependence on the energy in the intermediate state
is weak and the closure approximation is applicable
and one does not have to sum explicitly over the nuclear 
intermediate states.
Also, b) since $qR > 1$ ($R$ is the nuclear radius), the expansion
in multipoles is not convergent, unlike in the $2\nu$ decay.
In fact, all possible multipoles contribute by a comparable amount.
Finally, c) the neutrino propagator results in a neutrino potential
of a relatively long range (see Eq. (15)).

Thus, in order to evaluate the rate of the $0\nu$ decay, we need to
evaluate only the matrix element connecting the ground states $0^+$ of
the initial and final nuclei. Again, we can use the QRPA or the shell model.
Both calculations show that the features enumerated above are indeed
present. In addition, the QRPA typically shows less extreme dependence
on the particle-particle coupling constant $g_{pp}$
than for the $2\nu$ decay, since the contribution
of the $1^+$ multipole is relatively small. The calculations also
suggest that for quantitatively correct results one has
to treat the short range nucleon-nucleon repulsion carefully, 
despite the long range of the neutrino potential.

Does that mean that the calculated matrix elements are insensitive to
nuclear structure? An answer to that question has obviously great
importance, since unlike the $2\nu$ decay, we cannot directly test whether
the calculation is correct or not. 

\begin{figure}[h]
\includegraphics[width=.8\textwidth]{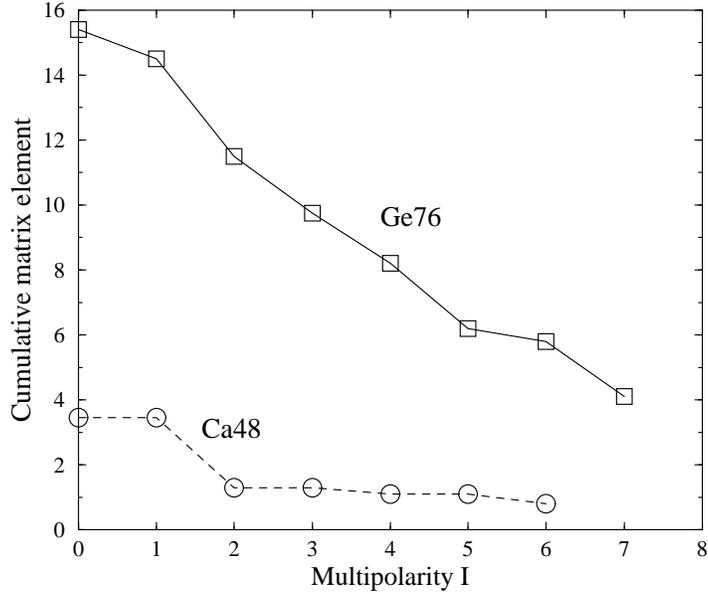}
\caption{The cumulative contribution,
i.e., the summed contribution of all natural
parity multipoles up to $I$ of the exchanged $nn$ and $pp$ pair, 
to the $0\nu$ nuclear matrix 
element combination $M_{GT}^{0\nu} - M_F^{0\nu}$.
The full line is for $^{76}$Ge and the dashed line for $^{48}$Ca.}
\label{fig:0nu}
\end{figure}

For simplicity, let us assume that the $0\nu$ $\beta\beta$ decay is
mediated only by the exchange of a light Majorana neutrino. The
relevant nuclear matrix element is then the combination 
$M_{GT}^{0\nu} - M_F^{0\nu}$, where the GT and F operators
change two neutrons into two protons, and contain the corresponding
operator plus the neutrino potential. One can express these matrix elements 
either in terms of the proton particle - neutron hole multipoles (i.e.,
the usual beta decay operators) or in terms of the multipole coupling
of the exchanged pair, $nn$ and $pp$. 

When using the decomposition in the proton particle - neutron hole multipoles,
one finds that all possible multipoles (given the one-nucleon states near
the Fermi level) contribute, and the contributions have typically equal signs.
Hence, there does not seem to be much cancellation.

However, perhaps more physical is the decomposition into the exchanged
pair multipoles. There one finds, first of all, that only 
natural parity multipoles
($\pi = (-1)^I$) contribute noticeably. And there is a rather severe
cancellation. The biggest contribution comes from the $0^+$
multipole, i.e., the pairing
part. All other multipoles, related to higher seniority states, 
contribute with an opposite sign. The final matrix element is then 
a difference of the pairing and higher multipole 
(or broken pair $\equiv$ higher seniority) parts, and
is considerably smaller than either of them. This is illustrated in 
Fig. \ref{fig:0nu} where the cumulative effect is shown, i.e., the
quantity
$M(I) = \sum_J^I \left[ M_{GT}^{0\nu}(J) - M_F^{0\nu}(J) \right]$
is displayed for $^{76}$Ge (from \cite{muto2}) and $^{48}$Ca
(from \cite{ca48sm}). Thus, the final result depends sensitively
on both the correct description of the pairing and on the admixtures
of higher seniority configurations in the corresponding initial
and final nuclei. It appears, moreover, that the final result
might depend on the size of the single particle space included.
That important question requires further study.

\begin{figure}[h]
\centerline{\rotate[r]{\epsfysize=10cm \epsfbox{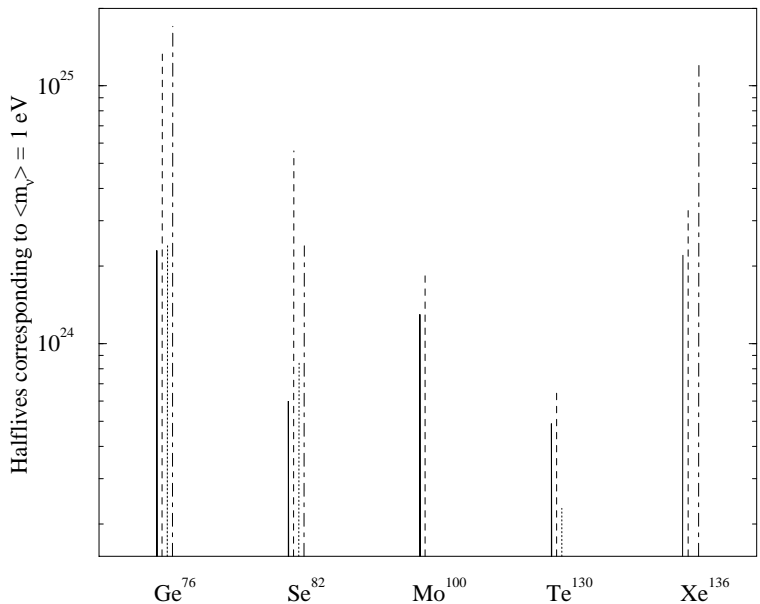}}}
\caption{ \protect
Half-lives (in years) calculated for $\langle m_{\nu} \rangle$ = 1 eV
by various representative methods and different authors
for the most popular double-beta decay candidate nuclei.
Solid lines are QRPA from \cite{Muto},
dashed lines are QRPA from \cite{Engel}
(recalculated for $g_A$ = 1.25 and $\alpha' = -390$~MeV fm${}^3$),
dotted lines are shell model \cite{HS},
and dot-and-dashed lines are shell model \cite{Caurier97}.
}
\label{fig:mini}
\end{figure}

Since there is no objective way to judge which calculation is
correct, one often uses the spread between the calculated 
values as a measure of the theoretical uncertainty.
This is illustrated in Fig.  \ref{fig:mini}. There, I have chosen
two representative QRPA sets of results, the highly truncated
``classical'' shell model result of Haxton and Stephenson,
and the result of more recent shell  model calculation which is 
convergent for the set of single particle states chosen
(essentially 0$\hbar \omega$ space). 

For the most important case of $^{76}$Ge the calculated rates
differ by a factor of 6-7. Since the effective neutrino mass
$\langle m_{\nu} \rangle$ is inversely proportional to the
square root of the lifetime, the experimental limit of
$1.6 \times 10^{25}$ y translates into limits of about 1 eV
using the NME of \cite{Engel,Caurier97}, and about 0.4 eV
with the NME of \cite{HS,Muto}. On the other hand, if one
would accept the more stringent limit of $5.7 \times 10^{25}$
\cite{Baud99}, even the more pessimistic matrix elements
restrict $\langle m_{\nu} \rangle < 0.5$ eV. 
Needless to say, a more objective measure of 
the theoretical uncertainty would be highly desirable. 
 
In Tables \ref{tab:0nu} and \ref{tab:maj} we list the deduced
limits on the fundamental parameters, 
the effective neutrino Majorana mass $\langle m_{\nu} \rangle$,
and the Majoron coupling constant $\langle g_{\nu,\chi} \rangle$.
The references to the source of the corresponding nuclear
matrix elements, used to translate the experimental half-life
limit into the listed limits on $\langle m_{\nu} \rangle$ and
$\langle g_{\nu,\chi} \rangle$ are also given.
When using the tables one has to keep in mind the uncertainties
illustrated in Fig. \ref{fig:mini}.

\subsection{Neutrinoless decay: very heavy Majorana neutrino}

\begin{figure}[h]
\begin{center}
\includegraphics[width=.5\textwidth]{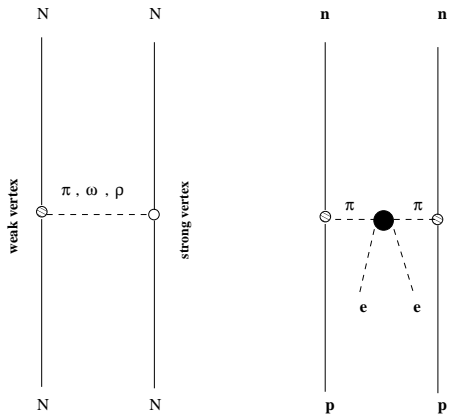}
\caption{The Feynman graph description of the parity-violating
nucleon-nucleon force (left graph) and of the \bb decay with
the exchange of a heavy neutrino mediated by the pion exchange.
The short range lepton number violating amplitude is symbolically
described by the filled blob in the right graph.}
\end{center}
\label{fig:moha}
\end{figure}

The neutrinoless $\beta\beta$ decay can be also mediated by the exchange
of a heavy neutrino. The decay rate is then inversely proportional 
to the square of the effective neutrino mass \cite{Verg81}. 
In this context it is particularly
interesting to consider the left-right symmetric model proposed by Mohapatra
\cite{Moh86}. In it, one can find a relation between 
the mass of the heavy neutrino $M_N$ and the mass of the right-handed 
vector boson $W_R$. Thus, the limit on the  $\beta\beta$ rate provides, 
within that specific model, a stringent
lower limit on the mass of $W_R$.

The process then involves the emission of the heavy $W_R^-$ by the first 
neutron and its virtual decay into an electron and the
heavy Majorana neutrino, $W_R^- \rightarrow e^- + \nu_N$.
This is followed
by the transition $\nu_N  \rightarrow e^- + W_R^+$ 
and the absorption of the $W_R^+$
on the second neutron, changing it into the
second proton. Since all exchanged particles between the two neutrons 
are very heavy, the corresponding ``neutrino potential'' is of essentially
zero range. Hence, when calculating the nuclear matrix
element, one has to take into account carefully the short range 
nucleon-nucleon repulsion.

As long as we treat the nucleus as an ensemble of nucleons only, the only
way to have nonvanishing nuclear matrix elements for the above process
is to treat the nucleons as finite size particles. 
In fact, that is the standard
way to approach the problem \cite{Verg81}; the nucleon size is described
by a dipole form factor with the cut-off parameter $\Lambda \simeq$  0.85 GeV.
Using such a treatment of the nucleon size, and the half-life limit for the
$^{76}$Ge $0\nu$ decay listed in Table \ref{tab:0nu}, one obtains
a very interesting limit on the mass of the vector boson $W_R$
\cite{Hir96}
\begin{equation}
m_{W_R}  \ge  1.6 {\rm ~ TeV}~.
\end{equation}

However, another way of treating the problem is possible, and already mentioned
in \cite{Verg81}. Let us recall how the analogous situation is treated
in the description of the parity-violating nucleon-nucleon force \cite{AH85}.
There, instead of the weak (i.e., very short range) interaction of two nucleons,
one assumes that a meson ($\pi, \omega, \rho$) is emitted by one nucleon
and absorbed by another one. One of the vertices is the parity-violating one,
and the other one is the usual parity-conserving strong one. 
The corresponding range is then
just the meson exchange range, easily treated. The situation is schematically
depicted in the left-hand panel of Fig. \ref{fig:moha}. 
The analogy for $\beta\beta$
decay is shown in the right-hand graph. It involves two pions, and the
``elementary'' lepton number violating
\bb decay then involves a transformation of two pions
into two electrons. Again, the range is just the pion exchange range.
It would be interesting to see if a detailed treatment of this
graph would lead to more or less
stringent limit on the mass of the $W_R$ than the treatment with form factors.
The relation to the claim in \cite{rho72} that an analogous graph
contributing to the lepton number violating muon capture identically vanishes
should be further investigated; in fact that claim is probably not valid.

\section{Perspectives}
As stated earlier, the present best limits on the rate of the $0\nu$ \bb decay,
or equivalently on the neutrino effective Majorana mass $\langle m_{\nu} \rangle$,
were obtained with an exposure of about 20 kg$\cdot$yr. Several experiments
(Heidelberg-Moscow $^{76}$Ge \cite{Baud99}, IGEX $^{76}$Ge \cite{IGEX99},
Caltech-Neuchatel TPC $^{136}$Xe \cite{Leusch98}) are  presently
at or near that level. The other limits in
Tables \ref{tab:0nu} and \ref{tab:maj} were obtained with 
smaller exposures of $\sim$ 1 kg$\cdot$yr. 
The detector NEMO-3, with planned source mass of 10 kg, is being built
and should be operational soon \cite{Nemo99}. In few years of operation
it should reach a half-life limit of $\sim 10^{25}$ yr for the $0\nu$ decay
of $^{100}$Mo, and perhaps other nuclei as well.
However, further improvements with the existing detectors 
becomes increasingly difficult, since the sensitivity 
to the $\langle m_{\nu} \rangle$ is proportional
to only the 1/4 power of the source mass and exposure time.
Thus, much larger sources are clearly needed. 

What are the perspectives of a radical improvement in the search for the
$0\nu$ \bb decay? To achieve that, one would have to build a detector 
capable of using hundreds of kg or even several tons of the source material.
At the same time, the background per unit mass has to be correspondingly
improved so that one can benefit from the larger mass. Obviously,
such program is very challenging. 

The difficulty begins with the problem
of acquiring  such a large mass of the isotopically separated and radioactively
clean material. Here, the principal obstacle is the cost of the isotope
separation. (This can be avoided only if the source isotope has large
abundance; in practice that is true only for $^{130}$Te with 34\% 
abundance.)

The second unavoidable difficulty is the background caused by the $2\nu$
decay. One can observe the $0\nu$ decay only if its
rate exceeds the fluctuations of the  $2\nu$ events
at the same energy, i.e., near the decay $Q$
value.
The number of $2\nu$ decays in an energy interval $\Delta E$ 
near $Q$ depends on these quantities like 
$\sim (\Delta E/Q)^6$ provided $\Delta E \ll Q$.  
(If the energy resolution is folded in, this dependence is somewhat modified.)
Thus, good energy resolution, which determines how wide
interval $\Delta E$ one must consider, is again crucial in order to 
reduce the effect of this ``ultimate'' background.

One of the proposal for such a large \bb experiments has been
extensively discussed in the literature (see, e.g., \cite{Kla99}).
The project, with the acronym GENIUS, would use a large amount
of `naked' enriched $^{76}$Ge, in the form of an
array of about 300 detectors, suspended in liquid nitrogen, 
which provides simultaneously cooling and shielding. 
It is envisioned that the detector would consist of one ton of 
enriched $^{76}$Ge. The anticipated background 
is 0.04 counts/(keV$\cdot$yr$\cdot$t), i.e., about 1000 times lower than the 
best existing backgrounds. Such a detector could reach the half-life
limit of about $6 \times 10^{27}$ yr within one year of operation,
thus improving the neutrino mass limit by an order of magnitude.

Another large project, CUORE, \cite{Fio98} is a cryogenic set-up
consisting of 17 towers, each containing 60 cubic crystals of TeO$_2$.
It would be housed in a single specially constructed dilution refrigerator
and would contain about 800 kg of the sensitive material. A prototype
system, CUORICINO, consisting of one of the towers, is being
developed now. 

An experiment with a large amount (100 tons) 
of natural molybdenum, (abundance of the \bb candidate 
$^{100}$Mo = 9.6\%)
with good energy and position resolution, is proposed
in Ref. \cite{Ejiri}.

In order to radically suppress the background, the  Ba ions,
the final products of the $^{136}$Xe double beta decay,
could be identified by laser tagging. That approach, described
in Ref. \cite{EXO}, would allow to use a large Time Projection
Chamber with perhaps ton quantities of  $^{136}$Xe, reaching
sensitivities to half-lives $\sim 10^{28}$ years.
 
This, still incomplete list of proposed very large \bb
decay experiments shows that the field is entering 
a critical phase.
If the new techniques, mentioned above, 
can be developed in conjuntion with the large
source mass, the background caused by radioactivities can be
essentially eliminated. However, as stated above, the ultimate
background due to the tail of the $2\nu$ decay can be compensated
only by a superior energy resolution.

Given the importance of the neutrinoless decay, it is likely that several
of these large and costly projects, involving ton$\cdot$years of exposure and 
a correspondingly reduced background, will be realized in the 
foreseeable future. Thus, sensitivity to the neutrino
Majorana mass $\langle m_{\nu} \rangle$ approaching
0.01 eV may be in sight. Whether the neutrinoless decay
will be discovered is unknown, but the reasons to look for
it are so compelling, that the search will undoubtedly
continue.

\section{Implications}

The study of  $0\nu$ \bb decay 
provides at present an upper limit well below 1 eV for the effective
electron neutrino Majorana mass
$\langle m_{\nu} \rangle$ even if the
most pessimistic nuclear matrix elements
are used. What are the consequences
of that limit when combined with the manifestations of the
neutrino oscillations? 

Recall that the
atmospheric neutrino anomaly (with its zenith angle
dependence)  implies the nearly maximum mixing
of $\mu$ and $\tau$ neutrinos (or $\mu$ and sterile
neutrinos) with $\Delta m^2 \sim 10^{-3}$
eV$^2$ (see chapter 5 of this book). 
There is, so far, no unique neutrino oscillation solution 
to the solar neutrino deficit(see chapter 4 of this book). 
However, all of the acceptable solutions
have $\Delta m^2 < 10^{-4}$ eV$^2$ and involve electron
neutrinos. Both large and small mixing angle solutions are
currently compatible with the data. 
Finally, the third ``positive'' evidence comes from the
LSND experiment (see chapter 7 of this book), 
and implies relatively small mixing
between the electron and muon neutrinos and
$\Delta m^2 \ge 0.1$ eV$^2$.
A full analysis must contain, in addition, all experimental results
which exclude various parts of the possible regions of the
quantities $\Delta m^2$ and the mixing angles. 
Taking all these findings together would necessarily imply
the existence of a fourth neutrino, which must be ``sterile''
given the constraint on the invisible width of the $Z$.
At the same time, it is well known that oscillation
experiments are not able to furnish the overall
scale of the neutrino masses.

This absolute neutrino mass scale is essential
not only as a matter of principle, but in particular
if one wants to  ascribe part of the dark matter,
namely its ``hot'' component, to massive neutrinos.
Doing that would mean that the
sum of the neutrino masses $\sum m_{\nu}$ is 
one or several eV. Tritium beta decay gives 
an upper limit of a similar magnitude for any
mass eigenstate with a large electron flavor component.
Clearly, if light neutrinos are responsible for 
a nonnegligible part of the dark matter, 
the oscillation data mean that at least two,
and possibly all neutrino masses are nearly
degenerate. (Such scenario was discussed for the first
time in Ref. \cite{David}.)  The relation of the $0\nu$
\bb decay and the oscillation scenarios, 
in particular the scenarios involving
degenerate neutrinos, has been a topic
of several recent papers 
\cite{GG98,EL99,BW99,Bil99}. 

The consequences are particularly 
dramatic if one assumes that
only three massive Majorana neutrinos
exist with nearly degenerate masses 
$m_i  \simeq \bar{m} \sim $ $O$(eV).
(Hence discarding for this purpose the LSND
experimental result, even though there is
no evidence against it.)
The \bb decay constraint can be expressed as \cite{GG98}
\begin{equation}
\langle m_{\nu} \rangle = | m_1 c_2^2 c_3^2 e^{i\phi}
+ m_2 c_2^2 s_3^2 e^{i\phi'} + m_3 s_2^2 e^{i2\delta} | ~,
\end{equation}
where $c_i, s_i$ denote $\cos \theta_i, \sin \theta_i$ in the
$3 \times 3$ mixing matrix, $\delta$ is the CP violating phase
in that matrix, and $\phi, \phi'$ are the CP violating phases
in the diagonal mass matrix.  Clearly, the differences
in $m_i$ can be neglected in this case. Moreover, the 
reactor long baseline experiments have
established that $\nu_e$ do not mix very much
with anything else near $\Delta m^2 \sim 10^{-3}$
eV$^2$, which means that 
the angle $\theta_2$ is small.
At the same time, the angle $\theta_1$ which controls
the atmospheric neutrino oscillations is near its
maximum value $\sin^2 2\theta_1 \simeq 1$.
Thus 
\begin{equation}
|\cos^2\theta_3 + \sin^2\theta_3 e^{i(\phi'-\phi)} | < 
\langle m_{\nu} \rangle/ \bar{m} \ll 1 ~.
\label{eq:gg}
\end{equation}
Hence also $\theta_3$ must be near the maximum
mixing, $\sin 2\theta_3 \simeq 1$, and the CP phases
in the above equation (\ref{eq:gg}) are such that
the two terms cancel each other.

That would be a very unexpected result. We would have three
massive highly degenerate neutrinos with bimaximal mixing.
Moreover, the electron neutrino would be `quasi-Dirac'
with its two components essentially canceling each other
in their contribution to the $0\nu$ \bb decay.
While such a scenario is rather problematic (see \cite{EL99}),
and  it does not accommodate the LSND result at all,
it illustrates the power of the neutrinoless
\bb decay in constraining the choice of the
neutrino oscillation scenarios.

\section{Conclusions}

The quest for neutrino mass is at a critical stage at present.
The evidence for neutrino mixing is getting
stronger and stronger, and the basic parameters
describing the neutrino oscillation phenomena
are being constrained more and more. At the same time,
the oscillation searches cannot give us the scale
of the neutrino masses, but only their differences.
Among the experiments that are sensitive to the
masses themselves, albeit to  their different
aspects (end point of the ordinary
beta decay, observation of the supernova
neutrinos, and the neutrinoless double beta decay),
only the $0\nu$ decay is able to reach the sub eV
region, and in a foreseeable future extend it by
a substantial margin.

In this review the present status of the \bb decay
is described.  The unpleasant uncertainty related
to the nuclear structure aspect of the problem
is estimated to be at
the level of a factor of 2-3 for the effective neutrino mass.
However, the experimental progress is such that even using
the most conservative nuclear matrix elements allows
us to push the limit well below the competing techniques.
The nuclear structure uncertainty can be reduced by
further development of the corresponding nuclear models.
At the same time, by reaching comparable 
experimental limits in several
nuclei, the chances of a severe error in the NME
will be substantially reduced.

Several projects are under way that will improve
the life-time limit substantially, or find the $0\nu$ decay.
Already now the search for \bb decay gives important
constraints on the fundamental properties of neutrinos
and their interactions. The role of the \bb decay in
the whole enterprise descibed in various chapters of this
book will be substantially
strengthened once these ambitious projects are underway.

\bigskip
{\it Acknowledgment}

This work was supported by the US Department of Energy
under Grant No. DE-FG03-88ER-40397.

%

\end{document}